\documentclass[acmsmall]{acmart}

\usepackage{microtype}
\usepackage[utf8]{inputenc}            
\usepackage[T1]{fontenc}               
\usepackage{mathpartir}                
\usepackage{mathwidth}                 
\usepackage{stmaryrd}                  
\usepackage{xspace}                    
\usepackage{hyperref}                  
\usepackage[scaled=0.85]{beramono}
\usepackage[customcolors,shade]{hf-tikz}
\usepackage{tikz}
\usetikzlibrary{calc}
\usetikzlibrary{arrows.meta}
\usetikzlibrary{decorations}
\hfsetfillcolor{gray!40}
\hfsetbordercolor{gray!40}
\usepackage{float}
\usepackage{multirow}
\usepackage{booktabs}

\usepackage{booktabs}   
\usepackage{subcaption} 

\usepackage{numdef}

\usepackage{listings}
\lstdefinestyle{wast}{
  belowcaptionskip=1\baselineskip,
  breaklines=true,
  frame=none,
  xleftmargin=2\parindent,
  language=lisp,
  showstringspaces=false,
  basicstyle=\sffamily\small,
  keywordstyle=\bfseries,
  commentstyle=\itshape,
  columns=fullflexible,
  keepspaces=true,
  escapechar=\#,
  morekeywords={barrier,bind,block,br,br_if,call,call_indirect,call_ref,cont,cont.bind,cont,else,new,func,handler,if,on,i32,i64,add,drop,const,eq,ge_u,gt_u,local,local.set,local.get,loop,param,ref,ref.func,result,resume,resume_throw,resume_with,return,suspend,suspend_to,tag,then,type,elem,declare,return_call,return_call_ref,eqz,f32,convert_i32_s,unreachable,ref.is_null,br_table}
}
\lstdefinestyle{rust}{
  belowcaptionskip=1\baselineskip,
  breaklines=true,
  frame=none,
  xleftmargin=2\parindent,
  language={},
  showstringspaces=false,
  basicstyle=\sffamily\small,
  keywordstyle=\bfseries,
  commentstyle=\itshape,
  columns=fullflexible,
  keepspaces=true,
  escapechar=\#,
  morekeywords={match, mut,let,trait,pub,fn,Self,self}
}
\lstdefinestyle{assembly}{
  belowcaptionskip=1\baselineskip,
  breaklines=true,
  frame=none,
  xleftmargin=2\parindent,
  language={},
  showstringspaces=false,
  basicstyle=\sffamily\small,
  keywordstyle=\bfseries,
  commentstyle=\itshape,
  columns=fullflexible,
  keepspaces=true,
  escapechar=\#,
  morekeywords={push,pop,mov,ret},
  morecomment=[l]{//},
}
\AtBeginDocument{%
  }

\setcopyright{rightsretained}
\acmPrice{}
\acmDOI{10.1145/3622814}
\acmYear{2023}
\copyrightyear{2023}
\acmSubmissionID{oopslab23main-p195-p}
\acmJournal{PACMPL}
\acmVolume{7}
\acmNumber{OOPSLA2}
\acmArticle{238}
\acmMonth{10}
\received{2023-04-14}
\received[accepted]{2023-08-27}

\newcommand\void[1]{}
\newcommand\simd[1]{}

\definecolor{hilite}{rgb}{0.7,0,0}
\newcommand*{\hilite}[1]{%
  \tikz[baseline=(X.base)] \node[rectangle, fill=gray!35, rounded corners, inner sep=0.3mm] (X) {#1};%
}

\newcommand\X[1]{\ensuremath{\mathit{#1}}}
\ifx\G\undefined
  \newcommand\G[1]{\ensuremath{\mathrm{#1}}}
\else
  \renewcommand\G[1]{\ensuremath{\mathrm{#1}}}
\fi
\newcommand\K[1]{\ensuremath{\mathsf{#1}}}
\newcommand\KK[1]{\textsf{\bfseries#1}}

\renewcommand\iff{\mathrel{\mbox{if}}}

\newcommand\production[1]{\mbox{(#1)}}

\numm
\numm
\numm
\numm
\numm

\numm
\numm
\numm
\numm

\numm
\numm
\numm
\numm
\numm

\numm
\numm

\numm
\numm
\numm\newcommand{\I32}{\K{i\scriptstyle32}}
\numm\newcommand{\I64}{\K{i\scriptstyle64}}

\numm
\numm
\numm

\newcommand\REF{\K{ref}}

\newcommand\CONT{\K{cont}}

\newcommand\valtype{\X{t}}
\newcommand\resulttype{\valtype^\ast}
\newcommand\functype{\X{f\!t}}

\newcommand\tagtype{\functype}

\newcommand\funcidx{\X{x}}

\newcommand\tagidx{\X{x}}

\newcommand\localidx{\X{x}}
\newcommand\labelidx{\X{l}}

\newcommand\MODULE{\KK{module}}
\newcommand\MFUNC{\KK{func}}

\newcommand\MTAG{\KK{tag}}

\newcommand\FLOCALS{\K{local}}

\newcommand\module{\X{m}}

\newcommand\func{\X{f}}
\newcommand\cont{\ensuremath{\mathit{cont}}}

\newcommand\tagg{\textit{tg}}

\newcommand\import{\X{im}}
\newcommand\export{\X{ex}}

\newcommand\BLOCK{\KK{block}}
\newcommand\LOOP{\KK{loop}}

\newcommand\BR{\KK{br}}
\newcommand\BRIF{\KK{br\_if}}

\newcommand\CALL{\KK{call}}
\newcommand\CALLREF{\KK{call\_ref}}

\newcommand\LOCALGET{\KK{local.get}}
\newcommand\LOCALSET{\KK{local.set}}

\newcommand\REFNULL{\KK{ref{.}null}}

\newcommand\REFFUNC{\KK{ref{.}func}}

\newcommand\CONST{\KK{const}}

\newcommand\CONTNEW{\KK{cont{.}new}}
\newcommand\CONTBIND{\KK{cont{.}bind}}
\newcommand\SUSPEND{\KK{suspend}}
\newcommand\RESUME{\KK{resume}}
\newcommand\RESUMETHROW{\KK{resume\_throw}}
\newcommand\THROW{\KK{throw}}

\newcommand\BARRIER{\KK{barrier}}
\newcommand\ON{\KK{on}}

\newcommand\blocktype{\functype}

\newcommand\instr{\X{e}}

\newcommand\handler{\X{h}}

\numm
\numm
\numm
\numm
\numm

\numm
\numm
\numm
\numm

\numm
\numm

\numm
\numm

\numm

\newcommand\textl{\mbox{‘}}
\newcommand\textr{\mbox{’}}
\renewcommand\text[1]{\textl\mathtt{#1}\textr}

\numm
\numm
\numm
\numm
\numm

\numm
\numm

\numm
\numm
\numm
\numm

\numm
\numm

\newcommand\ok{\mathrel{\mbox{ok}}}
\newcommand\okhandler{\mathrel{\mbox{clause}}}

\newcommand\CFUNC{\K{func}}
\newcommand\CTAG{\K{tag}}

\newcommand\CLOCAL{\K{local}}
\newcommand\CLABEL{\K{label}}
\newcommand\CRETURN{\K{return}}

\newcommand\vdashinstr{\vdash}
\newcommand\vdashinstrseq{\vdash}

\newcommand\vdashhandler{\vdash}

\newcommand\vdashfunc{\vdash}
\newcommand\vdashtag{\vdash}

\newcommand\vdashmodule{\vdash}

\newcommand\stepto{\hookrightarrow}

\newcommand\addr{\X{k}}

\newcommand\MIFUNCS{\K{func}}

\newcommand\MITAGS{\K{tag}}

\newcommand\SFUNCS{\K{func}}

\newcommand\STAGS{\K{tag}}

\newcommand\SCONTS{\K{cont}}

\newcommand\store{\X{s}}
\renewcommand\state{\X{z}}

\newcommand\LABEL{\KK{label}}
\newcommand\FRAME{\KK{frame}}

\newcommand\HANDLER{\KK{handler}}
\newcommand\blockargs[1]{\{#1\}}

\renewcommand\frame{\X{f\!\!f}}

\newcommand\REFCONTADDR{\KK{ref{.}cont}}
\newcommand\TRAP{\KK{trap}}

\newcommand\val{\X{v}}

\newcommand\config{\X{c}}

\newcommand{\ba}{\begin{array}}
\newcommand{\ea}{\end{array}}

\newenvironment{syntax}{\begin{displaymath}\ba{@{}l@{~\;}r@{~\;}c@{~\;}l@{}}}{\ea\end{displaymath}\ignorespacesafterend}
\newenvironment{reductions}{\begin{displaymath}\ba{@{}l@{~\;}c@{~\;}ll@{}}}{\ea\end{displaymath}\ignorespacesafterend}

\begin{document}

 \title{Continuing WebAssembly with Effect Handlers}

\author{Luna Phipps-Costin}
 \orcid{0009-0001-4398-6987}
 \email{phipps-costin.l@northeastern.edu}
 \affiliation{%
   \institution{Northeastern University}
   \city{Boston}
   \country{United States}
}

\author{Andreas Rossberg}
 \orcid{0000-0003-3137-3160}
 \email{rossberg@mpi-sws.org}
 \affiliation{%
   \institution{Independent}
   \city{Munich}
   \country{Germany}
}

\author{Arjun Guha}
\orcid{0000-0002-7493-3271}
\affiliation{%
 \institution{Northeastern University and Roblox}
 \country{United States}}
 \email{a.guha@northeastern.edu}

\author{Daan Leijen}
\orcid{0000-0003-1027-5430}
\affiliation{%
  \institution{Microsoft Research}
  \city{Redmond}
  \country{United States}}
\email{daan@microsoft.com}

\author{Daniel Hillerstr{\"o}m}
\orcid{0000-0003-4730-9315}
\affiliation{%
  \institution{Huawei Zurich Research Center}
  \country{Switzerland}}
\email{daniel.hillerstrom@ed.ac.uk}

\author{KC Sivaramakrishnan}
\orcid{0000-0002-3491-1780}
\affiliation{%
  \institution{Tarides and IIT Madras}
  \country{India}}
\email{kc@kcsrk.info}

\author{Matija Pretnar}
\orcid{0000-0001-7755-2303}
\affiliation{%
  \institution{University of Ljubljana and Institute of Mathematics, Physics \& Mechanics}
  \country{Slovenia}}
\email{matija.pretnar@fmf.uni-lj.si}

\author{Sam Lindley}
\orcid{0000-0002-1360-4714}
\affiliation{%
  \institution{The University of Edinburgh}
  \country{United Kingdom}
}
\email{sam.lindley@ed.ac.uk}

\renewcommand{\shortauthors}{Phipps-Costin, Rossberg, Guha, Leijen, Hillerstr{\"o}m, Sivaramakrishnan, Pretnar, and Lindley}

\begin{abstract}
  WebAssembly (Wasm) is a low-level portable code
  format offering near native performance.
  It is intended as a compilation target for a wide variety of source languages.
  However, Wasm provides no direct support for non-local control flow
  features such as async/await, generators/iterators, lightweight
  threads, first-class continuations, etc. This means that compilers
  for source languages with such features must ceremoniously transform
  whole source programs in order to target Wasm.

  We present \emph{WasmFX}, an extension to Wasm which provides a universal
  target for non-local control features via \emph{effect handlers},
  enabling compilers to translate such features directly into Wasm.
  Our extension is minimal and only adds three main
  instructions for creating, suspending, and resuming continuations.
  Moreover, our primitive instructions
  are type-safe providing typed continuations which are well-aligned with the
  design principles of Wasm whose stacks are typed.
  We present a formal specification of WasmFX and show that the
  extension is sound.  We have implemented WasmFX as an extension to
  the Wasm reference interpreter and also built a prototype WasmFX
  extension for Wasmtime, a production-grade Wasm engine, piggybacking
  on Wasmtime's existing fibers API.
  The preliminary performance results for our prototype are encouraging,
  and we outline future plans to realise a native implementation.
\end{abstract}




\maketitle

\section{Introduction}
\label{sec:intro}
WebAssembly (also known as\ Wasm)~\cite{Wasm, Wasm1, Wasm2} is a
low-level virtual machine designed to be safe and fast, while being
both language- and platform-independent.  A primary motivating use
case is efficient code execution on the Web, but Wasm is employed in
many other environments, such as edge and cloud computing, mobile,
embedded systems, and blockchains.  Due to its universal nature and
its mostly direct mapping to modern CPUs, Wasm is now being targeted
by a multitude of different languages.

However, Wasm currently lacks direct support for implementing
non-local control flow features appearing in many relevant languages,
such as generators/iterators, coroutines, futures/promises,
async/await, effect handlers, call/cc, and so forth.  For some
languages, such features are central to their identity or essential
for performance, e.g., to support massively scalable concurrency.  Of
course, it would be possible to extend Wasm with special support for
each individual feature, but that would be at odds with Wasm's
low-level spirit and does not scale to the next 700 non-local control
flow features.  But without native Wasm support, the only option often
left to users is a global transformation of source programs, which is
at odds with modularity and often inefficient.

We propose WasmFX, a unified structured mechanism that is sufficiently
general to cover present use-cases as well as being forwards
compatible with future use-cases, while also admitting efficient
implementations.  WasmFX mechanism is based on \emph{delimited
continuations} extended with multiple \emph{named control tags}
inspired by \citeauthor{PlotkinP13}'s effect
handlers~\cite{PlotkinP09,PlotkinP13}.  From an operational
perspective, we may view delimited continuations as the rest of a
computation from a particular point in its execution up to a
\emph{delimiter}.  From an implementation perspective, we can view
them as additional \emph{stacks} that execution can switch to.  Tags,
then, are an interface for the possible kinds of non-local transfers
of control that a computation may perform.  In addition to
versatility, effect handlers are supported by a decade's worth of
literature~\cite{effects-bibliography}, have a straightforward typing
discipline that aligns well with Wasm, and are proven to admit
efficient implementation
strategies~\cite{SivaramakrishnanDWKJM21,evidence_passing}.

Following an overview of Wasm and WasmFX (Section~\ref{sec:overview}), we make the following contributions:
\begin{itemize}
    \item a formal specification of a minimal extension of Wasm with typed continuations (Section~\ref{sec:language}),
    \item a number of applications displaying the versatility of the
      extended language (Section~\ref{sec:applications}),
    \item a prototype implementation based on the optimising Wasmtime
      compiler and a preliminary performance evaluation comparing it
      to the current current state of the art implementations of
      non-local control in Wasm (Section~\ref{sec:implementation}),
    \item a discussion of interesting deviations from the usual
      implementations of typed continuations, implementation concerns
      that influenced the current design, and potential alternative
      designs and extensions (Section~\ref{sec:discussion}).
\end{itemize}
We conclude by discussing related and future work.

\section{Overview}
\label{sec:overview}
\newcommand\topic[1]{\emph{#1.~}}

\subsection{A Short Wasm Primer}

Wasm defines a \emph{virtual instruction set architecture} that
closely mirrors the instruction sets common to modern CPUs.  Unlike
real hardware, however, Wasm is structured as a \emph{stack machine},
i.e., instead of named registers, instructions operate on a virtual
operand stack, a design that typically achieves a more compact code
representation.
For example, a sequence of instructions
\begin{lstlisting}
  (i32.const 30)  (i32.const 12)  (i32.add)
\end{lstlisting}
pushes two integers to the stack and then adds them, pushing the
result.  To keep sequences of instructions more readable, the Wasm
text format allows syntactic sugar in which operands are ``folded''
into an instruction.  For example, the above sequence can be written
as an expression:
\begin{lstlisting}
  (i32.add  (i32.const 30)  (i32.const 12))
\end{lstlisting}

\topic{Local variables} In addition, Wasm also provides \emph{locals},
which are virtual registers accessed via dedicated instructions
\lstinline|local.get| and \lstinline|local.set|.  For example, the of
local \lstinline|$i| can be incremented:
\begin{lstlisting}
  (local.set $i  (i32.add  (local.get $i)  (i32.const 1)))
\end{lstlisting}

\topic{Blocks} In contrast to most other low-level code formats,
control flow in Wasm is \emph{structured}: it does not have arbitrary
goto, but merely outward branches to the end of a surrounding
\lstinline|block| (like \texttt{break} in C) or to the beginning of a
\lstinline|loop| (like \texttt{continue} in C).  In order to jump out
of a loop, we nest the loop immediately inside a block, for example
\begin{lstlisting}
  (block $b
     (loop $l
        (br_if $b  (i32.ge_u  (local.get $i)  (i32.const 42)))
        (local.set $i  (i32.add  (local.get $i)  (i32.const 1)))
        (br $l) ) )
\end{lstlisting}
increments the local \lstinline|$i| in a loop.  For each iteration, it
checks if \lstinline|$i| has exceeded 42, in which case it branches to
the end of the block \lstinline|$b|.  Otherwise, it proceeds with
incrementing \lstinline|$i| and repeats by branching to the beginning
of loop \lstinline|$l|.

\topic{Functions}
Wasm code is organised into functions.
Besides parameter and result types, functions may declare locals.
For example, the function \lstinline|$range| processes a given range of sequential integers:
\begin{lstlisting}
(func $range (param $from i32) (param $to i32)  (local $i i32)
   (local.set $i  (local.get $from))
   (block $b
      (loop $l
         (br_if $b  (i32.gt_u  (local.get $i)  (local.get $to)))
         (call $process  (local.get $i))
         (local.set $i  (i32.add  (local.get $i)  (i32.const 1)))
         (br $l) ) ) )
\end{lstlisting}
It declares a local variable \lstinline|$i|, initialising it to the parameter \lstinline|$from| and iterating up to \lstinline|$to|, processing each value using a previously defined function \lstinline|$process| which
just prints its argument,
\begin{lstlisting}
  (func $process (param $x i32)
    (call $print  (local.get $x))
\end{lstlisting}
such that the following call prints ``\lstinline|10 11 12 13|'':
\begin{lstlisting}
  (call $range  (i32.const 10)  (i32.const 13))
\end{lstlisting}

\topic{Function references} Wasm 2.0 adds simple \emph{reference
types}, which represent first-class pointers to functions or host
resources~\cite{Wasm2}.  References are opaque values, such that their
actual representation cannot be inspected and their use cannot
compromise memory or type safety.  The instruction
\lstinline|ref.func $f| produces a reference to a previously defined
function \lstinline|$f|, whilst \lstinline|call_ref| calls a function
through such a reference operand.  If we define \lstinline|$task| to
be the type of functions that takes a single \lstinline|i32|
parameter, we can write a function \lstinline|$run| that calls two
such functions sequentially:
\begin{lstlisting}
(type $task (func (param i32)))
(func $run (param $task1 (ref $task)) (param $task2 (ref $task))
   (call_ref $task  (i32.const 10)  (local.get $task1))
   (call_ref $task  (i32.const 20)  (local.get $task2)) )
\end{lstlisting}
We can define two functions that print given ranges of numbers and pass their references to \lstinline|$run|:
\begin{lstlisting}
(func $task1 (param $x i32)  (call $range  (local.get $x)  (i32.const 13)))
(func $task2 (param $x i32)  (call $range  (local.get $x)  (i32.const 23)))
(func $main  (call $run  (ref.func $task1)  (ref.func $task2)))
\end{lstlisting}
Executing \lstinline|$main| prints ``\lstinline|10 11 12 13 20 21 22 23|''.

\subsection{Continuations}

Let us now turn from plain Wasm to WasmFX.  A \emph{continuation} is a
first-class program object representing the remaining computation from
a certain point in the execution of a program.  WasmFX is based on a
structured notion of a \emph{delimited continuation}, which is a
continuation whose extent is delimited by some \emph{control
delimiter}, meaning it represents the remaining computation from a
certain point up to (and possibly including) its delimiter.
Intuitively, a continuation represents the current stack, whereas a
delimited continuation represents a segment of the stack.  In
implementations, stack segments are often realised by allocating
\emph{multiple} stacks in a dynamic parent-child hierarchy.

WasmFX introduces three core instructions for manipulating delimited
continuations.  First, \lstinline|cont.new| creates a new continuation
from a given function.  At the implementation level, this can be
thought of as \emph{creating a new stack}.  Next, \lstinline|suspend|
suspends the currently running computation and reifies it into a
continuation.  In terms of implementation, this can be viewed as
\emph{switching} from the current stack to its parent.  Finally,
\lstinline|resume| invokes a given continuation in a delimited scope
and declares how to handle any suspensions that happen inside it.  In
an implementation, that means switching from a parent stack to a
selected child.
For example, the code
\begin{lstlisting}
  (type $cont (cont $task))
  (cont.new $cont  (ref.func $task1))
\end{lstlisting}
declares the type \lstinline|$cont| to be continuations of the
function type \lstinline|$task|, and creates a continuation that
represents a suspended computation to print a sequence of integers.

\subsection{Tags}

A running computation suspends itself by invoking a declared
\emph{control tag}, which transfers control to the nearest
\emph{handler} for that tag (Section~\ref{sec:handlers}).
This way, control tags are similar to exceptions, with the key
difference being that execution may subsequently be resumed by the
handling context.  Tags, however, are not only resumable
exceptions~\citep{Steele90}, as handlers not only return a result to
the suspended computation, but can control when and how it is
resumed.

For example, let us declare a tag \lstinline|$yield| that allows a
current computation to signal (e.g. to a scheduler) that it is ready
to relinquish control of execution. We can then extend the function
\lstinline|$process| to yield after printing out the given number:
\begin{lstlisting}
  (tag $yield)
  (func $process (param $x i32)
     (call $print  (local.get $x))
     (suspend $yield) )
\end{lstlisting}
However, \lstinline|suspend| only creates a continuation at the
invocation site of the control tag, it does not determine how to
handle it (in contrast to delimited continuations, which do both).
Thus, executing
\begin{lstlisting}
  (call $run  (ref.func $task2)  (ref.func $task1))
\end{lstlisting}
prints \lstinline|0| and then traps, as there is no handler associated with \lstinline|$yield|.

\subsection{Handlers}
\label{sec:handlers}

How to react to \lstinline|$yield| is specified by installing a
suitable \emph{handler}, which may be done each time a continuation is
resumed. A handler both delimits the scope of continuations and
determines the behaviour of subsequent control suspensions inside
them.  The \lstinline|resume| instruction defines a handler in terms
of a jump table pairing control tags with labels pointing to
corresponding handler code.

Consider the following example, which treats \lstinline|$yield| as a
non-resumable exception. We begin by setting a local variable
\lstinline|$k| to store a reference to the continuation.
\begin{lstlisting}
  (local $k (ref $cont))
  (cont.new $cont  (ref.func $task1))
  (local.set $k)
\end{lstlisting}

Next, we start an encompassing block \lstinline|$h|, whose label
serves as a join point. Immediately inside \lstinline|$h|, we define
another block with label \lstinline|$on_yield|, which will serve as a
label for the handling code. Recall that in Wasm, branching to a block
continues execution \emph{after} the block, thus the code used to
handle \lstinline|$yield| is that immediately after the
\lstinline|$on_yield| block. Handler code receives the suspended
continuation on the stack, thus the block is annotated with a
\lstinline|ref $cont| result type.
\begin{lstlisting}
  (block $h
     (block $on_yield (result (ref $cont))
        (resume $cont (on $yield $on_yield)  (local.get $k))
        (call $print  (i32.const -2))
        (br $h) )
     ;; $on_yield lands here
     (call $print  (i32.const -1))
     (br $h) )
\end{lstlisting}
Having set both blocks up, we may resume the continuation using
\lstinline|resume|, which points to the handling code for
\lstinline|$yield| and invokes the continuation referenced by
\lstinline|$k|. The code that follows the resumption describes
printing -2, but will never execute since the continuation will
suspend with \lstinline|$yield|. Thus the program prints -1 and
exits. The ultimate output is the number 10, printed by
\lstinline|$task1|, followed by -1, printed by the handling code.

The \lstinline|resume| instruction pushes the continuation created by
\lstinline|(suspend $yield)| to the stack before branching to a
handler. For example, we may store it back into \lstinline|$k| and
repeat the process by introducing a loop \lstinline|$l|.
\begin{lstlisting}
  (block $h
     (loop $l
        (block $on_yield (result (ref $cont))
           (resume $cont (on $yield $on_yield)  (local.get $k))
           (call $print  (i32.const -2))
           (br $h) )
        ;; $on_yield lands here, with the continuation on the stack
        (local.set $k)  ;; grab continuation and save it
        (call $print  (i32.const -1))
        (br $l) ) )
\end{lstlisting}
This then alternates between printing numbers from \lstinline|$task1|
and printing -1, ultimately printing -2 once the continuation runs to
its end.  In terms of implementation, this can be thought of as
switching back and forth between the main stack and the stack for
\lstinline|$task1|.

The \lstinline|resume| instruction consumes its continuation operand,
meaning a continuation may be resumed only once --- i.e., we only
support \emph{single-shot} continuations.  An attempt to resume the
same continuation again will result in a trap (an extension to support
multi-shot continuations would be interesting, but difficult to
support efficiently or robustly in existing Wasm engines).
When a continuation runs to completion, i.e., when control is
transferred back to \lstinline|resume| via a return rather than a
suspension, then no further continuation is created.  The underlying
child stack can be considered dead and be reclaimed by the system.

\subsection{Using Continuations for Scheduling}

Handlers give us fine control over how to resume continuations. For
example, we can use them to implement lightweight threads (a primary
use-case for handlers). Let us assume a suitable interface for a queue
implementation in which we keep inactive threads.
While the queue is non-empty, the scheduler repeatedly resumes the
thread at the head of the queue, by way of a loop \lstinline|$l|
containing a block \lstinline|$on_yield| in which the continuation is
resumed. On any \lstinline|$yield| suspension, we branch to the
instructions following the block, which enqueue the current
continuation and repeat the process:
\begin{lstlisting}
   (block $scheduler
     (loop $l
       (br_if $scheduler (call $queue_empty)) ;; exit if the queue is empty
       (block $on_yield (result (ref $cont))
         (resume $cont (on $yield $on_yield)  (call $dequeue))
         ;; we land here when a thread finishes without suspensions
         (br $l) ) ;; repeat the loop and proceed with a smaller queue
       ;; we land here when a thread suspends with $yield, with continuation on stack
       (call $enqueue)
       (br $l) ) )
\end{lstlisting}
Now, enqueuing tasks \lstinline|$task1| and \lstinline|$task2| as
before and running the scheduler, we get interleaved output:
``\lstinline|10 20 11 21 12 22 23 33|''.

\void{ Instead of using a loop, the scheduler can also be implemented
  in a more functional style with a recursive function taking the
  active thread as an argument. For simplicity, we retain the inactive
  threads in a global queue.
\begin{lstlisting}
(func $scheduler (param $nextk (ref $cont))
  (if  (ref.is_null  (local.get $nextk))  (then (return)))
  (block $on_yield (result (ref $cont))
    (resume $cont (on $yield $on_yield)  (local.get $nextk))
    (return_call $scheduler  (call $dequeue)) )
  (call $enqueue)
  (return_call $scheduler  (call $dequeue)) )
\end{lstlisting}
As before, we resume the next thread, except that we obtain it from
the parameter rather the beginning of the queue. If the thread is
suspended with \lstinline|$yield|, we enqueue the suspended
continuation and reinvoke the function with the head of the queue. In
case the queue is empty, the function is supplied with a null
reference that causes it to return. To avoid this blowing the stack,
we use the \lstinline|return_call| instruction, which implements a
tail call~\cite{WasmTailcallProposal}.}

\section{Language}
\label{sec:language}
\subsection{Syntax}

\begin{figure}[b]
  \input{figures/fig-syntax}
  \caption{WasmFX Syntax}
  \label{fig:syntax}
\end{figure}

In the formal definition in this paper, we focus on a minimal language
\emph{WasmFX}, given in Figure~\ref{fig:syntax}, and omit much of
Wasm's complete feature set~\cite{Wasm2}, since it is largely
orthogonal to our proposal (features omitted relative to Wasm 2.0
include additional control instructions, arithmetics, floating point
and vector types, subtyping, globals, tables, and linear
memory). Nevertheless, the proposal is compatible with the full
language, and in examples we take the liberty to assume a more
complete instruction set, e.g., obvious features such as arithmetics.

Our proposal builds on top of extensions of Wasm with function
references~\cite{WasmFuncrefProposal}, which we extend to
continuations, and with exceptions~\cite{WasmExnProposal}, from which
we adapt \emph{tag} declarations. We also use exceptions whenever we
want to abort a continuation that we do not intend to resume. Both of
these extensions are not yet in the standard, but part of proposals
that are close to reaching the final states of standardisation.
Features added by our proposal are highlighted in \hilite{grey} in our
figures. But let us turn to the pre-existing constructs (in black)
first.

\topic{Types} As a low-level language, Wasm only provides primitive
data types available in hardware, such as integer numbers of different
bit width, such as $\I32$ and $\I64$. To support function references,
Wasm 2.0 introduced \emph{reference types} of the form
$\REF\functype$, where a \emph{function type} is of the form
$\valtype_1^\ast \to \valtype_2^\ast$, mapping arguments
$\valtype_1^\ast$ to results $\valtype_2^\ast$. In our extension,
continuations are assigned a similar type, except annotated with a
$\CONT$ keyword.

\topic{Instructions} The core building block of Wasm are
\emph{instructions}. The instruction set we consider here includes
basic constants, locals, control blocks $\BLOCK$ and $\LOOP$, and
basic branches $\BR$ and $\BRIF$. Certain instructions are annotated
with types (type-set as subscripts) to ensure unique typing. In
examples, we denote locals, functions, tags, or labels through
symbolic names of the form \lstinline|$name|, but these just stand in
for integer de~Bruijn indices that index into respective definition
lists.

We include $\CALLREF$~\cite{WasmFuncrefProposal}, which performs a
function call through a first-class function, a null reference
$\REFNULL$, and a function reference $\REFFUNC$. When calling a
reference, a runtime check that it is not null occurs; if it is,
execution \emph{traps}, i.e. immediately aborts. Given $\CALLREF$, a
regular $\CALL\;x$ instruction invoking function $x$, as available in
bare Wasm, can be viewed as a shorthand for the instruction sequence
$(\REFFUNC\,x)\,(\CALLREF_{\!\functype})$, so we omit it from our
core. After declaring an exception tag $x$, we may throw it using the
$\THROW\,x$ instruction. The exception proposal~\cite{WasmExnProposal}
also contains an instruction for handling exceptions, which we omit
for simplicity.

Next, the new instructions! As described in
Section~\ref{sec:overview}, $\CONTNEW_{\!\functype}$ creates a new
continuation (i.e., a stack) from a function reference,
$\RESUME_{\!\functype}\,h^\ast$ resumes it under the specified
handlers $h^\ast$, and $\SUSPEND\,x$ suspends the current continuation
and transfers control to the handler for tag $x$.

WasmFX adds two more instructions for special purposes.  The
instruction $\CONTBIND$ partially applies a given continuation. Unlike
partial application for function references, this instruction requires
no allocation: since a continuation can be resumed only once, an
implementation can modify it in-place.
To abort and finalise a continuation, we reuse the existing exception
mechanism: the $\RESUMETHROW\,x\,h^\ast$ instruction injects an
exception $x$ at the suspension site and thereby unwinds the aborted
stack.  Correct code is expected to consume all continuations
\emph{linearly}, i.e., either resume them or abort them explicitly
with $\RESUMETHROW$ --- both will ultimately terminate the computation
and reclaim the stack.  This prevents memory leaks and triggers any
user defined resource finalisation that was in place when the
continuation was suspended (e.g., through exception
handlers). Handling exception $x$ may trigger further suspensions,
which are handled with $h^\ast$.


\topic{Modules} All Wasm code is organised into modules. Each module
encapsulates local definitions, such as functions or tags, which can
optionally be exported or imported from other modules. Here, we only
include the bare minimum of concepts from modules. In particular, we
assume a single global module and omit the ability to export or import
definitions, which is needed to link multiple modules together. In
particular, the full language allows importing/exporting tags.

A tag definition includes a function type whose argument types
determine the arguments of the exception. For an exception, the result
of this function type must be empty, but when used as control tags,
the result types specify values to be transferred back upon
resumption.

\subsection{Typing}

Wasm code is typed in a context $C$ recording types of locals,
functions, tags and labels. Recall that Wasm uses de Bruijn indices,
thus contexts are simply indexed lists for each namespace. We denote
the type associated with label $l$ by $C_\CLABEL(l)$, and similarly
for other constructs.
\begin{syntax}
  \production{contexts} & C &::=&
    \epsilon ~|~
    C,\CFUNC~\functype ~|~
    C,\CTAG~\tagtype ~|~
    C,\CLOCAL~\valtype ~|~
    C,\CLABEL~\resulttype
\end{syntax}
To each Wasm instruction we assign a function type through a judgement
$C \vdash e : t_1^\ast \to t_2^\ast$ describing the types $t_1^\ast$
of values it expects to pop off the stack and the types $t_2^\ast$ of
values it is expected to push back. The typing rules for instructions
are given in Figure~\ref{fig:typing-instructions}.

\begin{figure}[h]
  \input{figures/fig-typing-instructions}
  \input{figures/fig-typing-rest}
  \caption{Typing Rules for WasmFX}
  \label{fig:typing-instructions}
\end{figure}

\void{Instruction sequences must have matching types. The third rule
  is a form of ``frame'' rule that allows instructions to be run on a
  larger stack.}
The first half of the figure contains the typing rules for plain Wasm
instructions, essentially unchanged from \citet{Wasm}.
We add the rules for $\CALLREF$, $\REFNULL$, $\REFFUNC$ and $\THROW$
taken from aforementioned proposals, which should be
self-explanatory.%
\void{First, the constant instruction does not pop off anything off
  the stack, but pushes back a value of the associated type. A similar
  condition holds for reading a local, while setting a local consumes
  the value on top of the stack and pushes nothing back. Since the
  expression $\instr^\ast$ inside $\BLOCK_\blocktype~\instr^\ast$ can
  run uninterrupted, the two need to have the same type. However, in
  the context, we associate the result type $t_2^\ast$ to the label,
  as this is what is expected on top of the stack after a branch. The
  rule for $\LOOP_\blocktype~\instr^\ast$ is similar, except that we
  expect the result type to be $t_1^\ast$, since $\instr^\ast$ will be
  executed after any branch. The typing rule for the branch
  instruction $\BR~l$ ensures that the expected values of type
  $t^\ast$ assigned to $l$ are on top of the stack, while the rest is
  not important, since the execution will be interrupted anyway. The
  conditional branch $\BRIF~l$ expects a conditional value of type
  $\I32$ in addition to the result of type $t^\ast$. Since in this
  case, the branch is conditional, there must be no other values on
  the stack, and any potential subsequent instructions can expect a
  stack of the same type.

  Next, the $\CALLREF$ instruction expects a function reference of
  type $\functype$ on top of the stack, and corresponding arguments,
  and pushes back the results of the function call. The types of
  $\REFNULL$ and $\REFFUNC$ are as expected and reflect the fact that
  a new reference of an appropriate type is pushed onto the stack.
}%
\footnote{The function references proposal~\cite{WasmFuncrefProposal}
distinguishes non-nullable references, but we omit that distinction.}
\void{Finally, the rule for $\THROW$ is similar to the one for
  $\BR$. We expect to find tag arguments on top of the stack, and
  ignore the rest since the execution will be interrupted.}

Note that $\BR$ and $\THROW$ are \emph{stack-polymorphic}: they allow
arbitrary inputs $t_1^\ast$ and outputs $t_2^\ast$ to be assumed for
the stack --- that is fine, because these instructions pass control
unconditionally and never return.

Let's turn our attention to what's new. The $\CONTNEW$ instruction
converts a function reference into a continuation reference of
analogous type. The $\RESUME$ instruction expects a continuation of
type $t_1^\ast \to t_2^\ast$ on the stack. The stack must also hold
values of types $t_1^\ast$ to be consumed by the continuation. If the
continuation terminates and hence $\RESUME$ returns, it yields results
of type $t_2^\ast$. The handler clauses are assigned types matching
the result type of the continuation ($t_2^\ast$). The auxiliary
judgement $C \vdashhandler \ON~\tagidx~\labelidx : t_2^\ast
\okhandler$ defined at the end ensures that each label $\labelidx$
used to handle a tag $\tagidx$ expects the corresponding tag arguments
${t'_1}^\ast$ and a continuation reference as parameters. The
continuation will produce $t_2^\ast$ (corresponding to the type of the
handler) once supplied with the tag results ${t'_2}^\ast$.  Typing of
$\RESUMETHROW$ is analogous, except that it expects arguments for the
thrown tag~$x$ instead of the continuation parameters.

The $\SUSPEND$ instruction behaves similarly to $\THROW$ in that it
expects appropriate tag operands of type $t_1^\ast$. The main
difference is that the continuation can be subsequently resumed with
values of type $t_2^\ast$. The $\CONTBIND$ instruction consumes and
produces a continuation reference, which is the same except that its
leading arguments $t^\ast$ have been partially applied.
\void{ Like $\BLOCK$ or $\LOOP$, the $\BARRIER$ instruction is not
  expected to change the behaviour of its body unless an unhandled tag
  is suspended, thus its type is the same as that of its body.  }

Typing rules for the included module-level constructs are
straightforward. When we declare a function of a type $t_1^\ast \to
t_2^\ast$, we may declare locals of type $t^\ast$ in addition to the
locals of type $t_1^\ast$ that hold the function's arguments.  The
context $C$ for typing a module is constructed recursively from the
types of the individual definitions. In effect, a module is one big
recursive definition.

\subsection{Execution}

The operational semantics of Wasm is given in terms of small-step
reductions between configurations consisting of the executed
instruction sequence together with a global store and the current
stack frame. To concisely express the reductions, we introduce
additional syntax in Figure~\ref{fig:reduction-syntax}.  Again, it is
largely inherited from plain Wasm~\cite{Wasm,Wasm2}; we focus on the
novelties.

\begin{figure}
  \input{figures/fig-reduction-syntax}
  \caption{Syntax of WasmFX Reductions}
  \label{fig:reduction-syntax}
\end{figure}

Instructions are extended into \emph{administrative instructions},
which represent internal values or operators not directly expressible
in Wasm's source instruction set.  The new instruction
$\REFCONTADDR~\addr$, which also is a value, represents a reference to
a continuation allocated at address $\addr$ in the global store.  The
instruction $\HANDLER\blockargs{\handler^\ast}~\instr^\ast$ represents
execution of $\instr^\ast$ under an active handler~$\handler^\ast$.
Handlers can occur as part of evaluation contexts.
For specifying the semantics of suspension, we use a restricted
version of \emph{handler contexts}~$E^\tagidx$, which are similar to
evaluation contexts, except that they cannot contain any handlers for
the given tag index~$\tagidx$.
The store holds function and tag declarations as well as allocated
continuations. A frame consists of a sequence of local values. A
continuation is given either by an evaluation context, or by the
sentinel token $\K{dead}$ when it is already used up.

\begin{figure}
  \input{figures/fig-reduction}
  \caption{WasmFX Reduction}
  \label{fig:reduction}
\end{figure}

The reduction rules are given in Figure~\ref{fig:reduction} and are of
the form $\store_1; \frame_1; \instr_1^\ast \stepto \store_2;
\frame_2; \instr_2^\ast$. For brevity, we omit the store when it is
not used; the same convention applies to frames.  We write $\frame[x =
  \val]$ to denote the same frame as $\frame$ but with the value
$\val$ assigned to local $x$.  Similarly, $\store[\addr = E]$ binds
the continuation $k$ to $E$ in store $\store$, and $\store[+\addr =
  E]$ is the \emph{extension} of $\store$ with a fresh continuation
$k$.

The first set of rules involving locals, blocks, labels and branches,
is once more inherited from plain Wasm, and we refer to \citet{Wasm}
for details.  The only change here is that we allow branches to cross
handlers, which uninstalls the handler.  A call to a function
introduces a suitable frame encapsulating the function
body. Evaluation is performed inside a frame until it either results
in values or throws.

The $\CONTNEW$ instruction extends the store with a fresh continuation
that will immediately invoke the referenced function when it is
resumed.

A continuation is resumed by installing a handler and inside it,
reestablishing the evaluation context that represents the
continuation. At the same time, the continuation is marked dead, so it
cannot be resumed twice. A handler is removed once evaluation
terminates with values. If it suspends with a tag for the handler, and
there is no intervening handler for it in-between (expressed by the
$E^\tagidx$ context), then the appropriate number of tag arguments are
extracted and the rest of the evaluation context up to the handler is
reified into a freshly allocated continuation, arguments and
continuation are pushed to the stack, and execution branches to the
label associated with the tag.

Partially applying a continuation with $\CONTBIND$ discards the old
continuation in the store and adds a new one with the evaluation
context extended with the supplied arguments.

Executing $\RESUMETHROW$ also installs a handler, reestablishes the
continuation's evaluation context, and marks the continuation
dead. However, it inserts a $\THROW$ instruction into the hole.

Finally, any attempt to consume a continuation that is already dead
causes a trap.

\subsection{Soundness}

Type Preservation for Wasm~\cite{Wasm,WasmCert} extends to the new
instruction set.  To prove this, we need to formulate adequate typing
rules for the new administrative instructions:
\begin{mathpar}
\infer{
  \store \vdash \store_\SCONTS(\addr) : t_1^\ast \to t_2^\ast
}{
  \store; C \vdashinstr \REFCONTADDR~\addr : \epsilon \to (\REF~\CONT~t_1^\ast \to t_2^\ast)
}
\and
\infer{
  (C^\circ \vdashhandler \handler : t^\ast \okhandler)^\ast
  \qquad
  C^\circ \vdashinstrseq \instr^\ast : \epsilon \to t^\ast
}{
  C \vdashinstr \HANDLER\blockargs{\handler^\ast}\, \instr^\ast : \epsilon \to t^\ast
}
\end{mathpar}
Here, $C^\circ$ is the same as $C$ but with all bindings for locals
and labels removed.  This restriction is necessary, because Wasm does
not have closures.  Without this restriction, we would allow dangling
occurrences of local or label names to enter the store in the
reduction rule for $\SUSPEND$.  The restriction is preserved because
all continuations start, via $\CONTNEW$, from a separate function.

With that, and a couple of additional auxiliary judgements not shown
here, we get:
\begin{theorem}[Preservation]
If $\vdash \store; \frame; \instr^\ast : t^\ast$
and $\store; \frame; \instr^\ast \stepto \store'; \frame'\!; {\instr'}^\ast$,
then $\vdash \store'; \frame'\!; {\instr'}^\ast : t^\ast$\!.
\end{theorem}

For Progress, we need to define what the legal \emph{results} of a
computation are, namely, either multiple values, a trap, or an
unhandled exception or suspension:
\begin{syntax}
   \production{result} & r &::=& \val^\ast ~|~ \TRAP ~|~ E[(\THROW~\tagidx)] ~|~ E^\tagidx[(\SUSPEND~\tagidx)] \\
\end{syntax}
Then we can prove, in the usual manner:
\begin{theorem}[Progress]
If $\vdash \store; \frame; \instr^\ast : t^\ast$,
then either $\instr^\ast = r$,
or $\store; \frame; \instr^\ast \stepto \store'; \frame'\!; {\instr'}^\ast$.
\end{theorem}

\section{Applications}
\label{sec:applications}
\subsection{Generators}

When using generators, we deal with two concurrent computations. One
sequentially consumes values while the other one produces (or yields)
them one at a time. We can capture the production of values with a
single tag \lstinline|$gen| which carries the produced value as a
parameter:

\begin{lstlisting}
  (tag $gen (param i32))
\end{lstlisting}
Then, we can write the following function which indefinitely produces
natural numbers from 0:

\begin{lstlisting}
  (func $naturals
     (local $n i32)  ;; zero-initialised
     (loop $l
        (suspend $gen  (local.get $n))
        (local.set $n  (i32.add  (local.get $n)  (i32.const 1)))
        (br $l) ) )
\end{lstlisting}

Produced values can be consumed with a handler for \lstinline|$gen|,
for example one that sums up values until one exceeds
\lstinline|$upto|:

\begin{lstlisting}
  (func $sum_until (param $upto i32) (result i32)
    (local $v i32)
    (local $sum i32)
    (local $k (ref $cont))
    (local.set $k  (cont.new  (ref.func $naturals)))
    (loop $l
       (block $on_yield (result i32 (ref $cont))
          (resume $cont (on $yield $on_yield)  (local.get $k))
          (unreachable)  ;; $naturals never returns, so we can never get here
       ) ;;   $on_yield (result i32 (ref $cont))
       (local.set $k)  ;; store the new continuation for later resumption
       (local.set $v)  ;; store the produced value
       (local.set $sum  (i32.add  (local.get $sum)  (local.get $v)))  ;; add $v to $sum
       (br_if $l  (i32.lt_u  (local.get $v)  (local.get $upto)))
    (local.get $sum) )
\end{lstlisting}

Now, although the function \lstinline|$naturals| is an infinite loop,
we get 5050 by calling:

\begin{lstlisting}
  (call $sum_until  (i32.const 101))
\end{lstlisting}

Using a global table of active generators, we can implement a more
ergonomic interface to generators, in which a programmer does not use
handlers directly, but interacts solely through functions
\lstinline|$init|, which takes a continuation and returns a unique
identifier, and \lstinline|$next|, which takes the identifier and
returns the next value.

\subsection{Dynamic Lightweight Threads}
\label{sec:dynamic-threads}

We can make our lightweight threads functionality from
Section~\ref{sec:overview} considerably more expressive by allowing
new threads to be forked dynamically. For that, we declare a new
\lstinline|$fork| tag that takes a continuation as a parameter and
(like \lstinline|$yield|) returns no result.

\begin{lstlisting}
  (tag $fork (param (ref $cont)))
\end{lstlisting}
Thus, instead of manually enqueuing the threads, we can use the
\lstinline|$fork| tag:

\begin{lstlisting}
  (suspend $fork  (cont.new $cont  (ref.func $task1)))
  (suspend $fork  (cont.new $cont  (ref.func $task2)))
\end{lstlisting}
Of course, we can obtain much more involved behaviour, as threads
themselves may fork new ones. The scheduler only needs to be extended
with a new clause for handling the \lstinline|$fork| tag:

\begin{lstlisting}
(func $scheduler (param $nextk (ref $cont))
   (loop $l
      (if  (ref.is_null (local.get $nextk))  (then (return)))
      (block $on_yield (result (ref $cont))
         (block $on_fork (result (ref $cont) (ref $cont))
            (resume $cont (on $yield $on_yield) (on $fork $on_fork)  (local.get $nextk))
            (local.set $nextk  (call $dequeue))
            (br $l) )
         ;; $on_fork, forkee and forker continuations on stack
         (local.set $nextk) ;; forker is next
         (call $enqueue)    ;; queue up forkee
         (br $l) )
         ;; $on_yield, yielder continuation on stack
      (call $enqueue)  ;; queue it up
      (local.set $nextk  (call $dequeue))  ;; take next
      (br $l) ) )
\end{lstlisting}

When handling a fork, there are two continuations on the stack. On
top, we have the suspended continuation, and below it, the newly
created thread that was given as an argument to \lstinline|$fork|. We
have three threads to choose from for the next step: the suspended
active thread, the newly created thread, and the first inactive thread
waiting in the queue. In the example above, we continue running the
active thread and enqueue the new one, but we could easily adopt a
different strategy.

\subsection{Promises}

We can adapt a similar approach for async/await-style
\emph{promises}~\cite{BiermanRMMT12}, in which one asynchronously runs
a function that will compute a value, obtains an opaque promise, and
then \emph{awaits} that promise once the value is required. It is
important to note that the function does not simply return the value,
but uses it to \emph{fulfill} a promise. If we represent promises with
$\I32$ identifiers, we can define a type of continuations that take
and fulfill them:

\begin{lstlisting}
  (type $i32_func (func (param i32)))
  (type $i32_cont (cont $i32_func))
\end{lstlisting}

We interface promises through four tags: \lstinline{$yield}
asynchronously yields as before, \lstinline{$fulfill} takes a promise
and a value to fulfill it with, \lstinline{$await} takes a promise,
awaits its fulfilled value, and returns it, and \lstinline{$async}
returns a new promise that will be fulfilled by the given
continuation. We represent values with $\I64$ in order to easily
distinguish them from promise identifiers.

\begin{lstlisting}
  (tag $yield)
  (tag $fulfill (param i32) (param i64))
  (tag $await (param i32) (result i64))
  (tag $async (param (ref $i32_cont)) (result i32))
\end{lstlisting}

Next, assume an external implementation of promises through externally
defined functions \lstinline{$prom_*} accessing a table that keeps a
(perhaps unfulfilled) value for each promise and a continuation that
awaits its result.
Then, we can write a handler similar to the one for light-weight
threads, except also handling the three additional tags. We describe
the body of the four handlers; the overall structure mirrors that of
the other examples.

When handling \lstinline{$fulfill}, the stack contains the suspended
continuation, the value, and the promise to be fulfilled. We keep the
first as the active continuation \lstinline{$nextk} and pass the other
two to the external implementation, which returns the continuation
awaiting the promise if any. If there is none, we do nothing,
otherwise, we enqueue it as it is now unblocked.

\begin{lstlisting}
  (local.set $nextk)
  (local.set $waiterk  (call $prom_fulfill))  ;; the call pops two operands off the stack
  (if  (i32.eqz  (ref.is_null  (local.get $waiterk)))   ;; i32.eqz encodes negation
     (then (call $enqueue  (local.get $waiterk))) )
\end{lstlisting}

When handling \lstinline{$await}, the stack contains the current
continuation and the promise it is waiting for. We first call the
external function and check if the promise has already been
fulfilled. If it has, then we can partially apply the current
continuation and resume it. Otherwise, we attach the continuation to
the promise and continue with the next thread in the queue.

\begin{lstlisting}
  (local.set $waiterk)  (local.set $prom)
  (if  (call $prom_fulfilled (local.get $prom))
     (then
        (local.set $nextk
            (cont.bind $i32_cont $cont  (call $prom_value  (local.get $prom))  (local.get $waiterk)) ))
     (else
        (call $prom_await  (local.get $prom)  (local.get $waiterk))
        (local.set $nextk  (call $dequeue)) ))
\end{lstlisting}

Finally, when handling \lstinline{$async}, the stack contains the
current continuation waiting for the new promise, and the asynchronous
task that is meant to fulfill it. We create a new promise and pass it
to both continuations; we enqueue the suspended continuation and run
the new one.

\begin{lstlisting}
  (local.set $waiterk)  (local.set $asynck)
  (local.set $prom  (call $prom_new))
  (call $enqueue  (cont.bind $i32_cont $cont  (local.get $prom)  (local.get $waiterk)))
  (local.set $nextk  (cont.bind $i32_cont $cont  (local.get $prom)  (local.get $asynck)))
\end{lstlisting}

\subsection{Actors}

As an example of handler composition, we consider Erlang-style
\emph{actors}~\cite{ArmstrongVWW96}. These are independent processes
that can spawn new actors and communicate to each other through
mailboxes. We represent their interface through four tags:
\lstinline{$send} sends a message to a given mailbox,
\lstinline{$receive} receives the next incoming message,
\lstinline{$spawn} creates a new actor from a given continuation and
returns its address, and \lstinline{$self} returns the address of the
current process. Like previously, we use $\I32$ for addresses and
$\I64$ for messages.

\begin{lstlisting}
   (tag $send (param i64 i32))
   (tag $self (result i32))
   (tag $spawn (param (ref $cont)) (result i32))
   (tag $recv (result i64))
\end{lstlisting}
To provide suitable type annotations, we need to define a number of
additional continuation types:
\begin{lstlisting}
  (type $i64_func (func (param i64)))   (type $i64_cont (cont $i64_func))
  (type $i64_cont_func (func (param i64 (ref $cont))))   (type $i64_cont_cont (cont $i64_cont_func))
  (type $cont_func (func (param (ref $cont))))   (type $cont_cont (cont $cont_func))
 \end{lstlisting}
As an example, here is a function that receives a message and forwards
it to actor \lstinline{$p}:
\begin{lstlisting}
  (func $forward (param $p i32)
     (local $s i64)
     (local.set $s  (suspend $recv))
     (suspend $send  (local.get $s)  (local.get $p)) )
\end{lstlisting}
And another that creates a chain of \lstinline{$n} actors that
forwards a message \lstinline{$m} to the originating actor:
\begin{lstlisting}
  (func $chain (param $n i32) (param $m i64)
     (local $p i32)  ;; mailbox of the currently last actor in the chain
     (local.set $p  (suspend $self))
     (loop $l
        (if  (i32.eqz  (local.get $n))
           (then  (suspend $send  (local.get $m)  (local.get $p)))  ;; once done, send $m to the last actor
           (else
              (cont.new $i32_cont  (ref.func $forward))  ;; set up a new forwarding continuation
              (cont.bind $i32_cont $cont  (local.get $p)  ;; pass it the address of the currently last actor
              (local.set $p  (suspend $spawn))  ;; spawn a new actor and set it as the currently last
              (local.set $n  (i32.sub  (local.get $n)  (i32.const 1)))  ;; decrement $n and repeat
              (br $l) ) ) ) ) )
\end{lstlisting}

If an actor tries to receive a message, but its mailbox is empty, it
must wait. The easiest way to implement this is by using threads as
described in Section~\ref{sec:dynamic-threads}. We assume an interface
to mailboxes through externally defined functions \lstinline{$mb_*}
and define a main function \lstinline{$act}, which takes a
continuation \lstinline{$k}, creates a new mailbox and passes its
address and the continuation to an auxiliary function
\lstinline{$act_aux}.
\begin{lstlisting}
  (func $act (param $k (ref $cont))
     (call $act_aux  (call $mb_new)  (local.get $k)))
\end{lstlisting}

The function \lstinline{$act_aux} stores the mailbox address in a
local \lstinline{$mine} and resumes the continuation under a handler
with the handler clauses as follows:

On \lstinline{$self}, we take the existing continuation expecting the
address, partially bind it to \lstinline{$mine}, and store it as the
next resumption.

\begin{lstlisting}
  (local.set $ik)
  (local.set $nextk  (cont.bind $i32_cont $cont  (local.get $mine)  (local.get $ik)))
\end{lstlisting}

On spawning a new actor \lstinline{$you}, we create a new mailbox,
store its address in \lstinline{$yours}, fork a new thread to run
\lstinline{$you} using \lstinline{$act_aux} again, and pass the new
address back to the continuation \lstinline{$ik}.

\begin{lstlisting}
  (local.set $ik)  (local.set $you)  (local.set $yours (call $mb_new))
  (suspend $fork
     (cont.bind $i64_cont_cont $cont  (local.get $yours)  (local.get $you)
        (cont.new $i64_cont_cont  (ref.func $act_aux)) ) )
  (local.set $nextk  (cont.bind $i32_cont $cont  (local.get $yours)  (local.get $ik)))
\end{lstlisting}

On sending a message, we simply pass the tag argument (message) to the
external function and resume:

\begin{lstlisting}
  (local.set $k)
  (call $mb_send)
  (local.set $nextk  (local.get $k))
\end{lstlisting}

Finally, on receiving a message, we first block until the mailbox
\lstinline{$mine} is no longer empty. Repeatedly, we check if it is
empty, and if it is, we yield the control to other actors and try
again later. Once the mailbox is non-empty, we read the message and
pass it to the continuation \lstinline{$ik}:

\begin{lstlisting}
  (local.set $ik)
  (loop $blocked
     (if  (call $mb_empty  (local.get $mine))
        (then  (suspend $yield)  (br $blocked)) ) )
  (local.set $nextk  (cont.bind $i64_cont $cont  (call $mb_recv  (local.get $mine))  (local.get $ik)))
\end{lstlisting}

Note that in the clauses for \lstinline{$spawn} and \lstinline{$recv},
we use tags \lstinline{$yield} and \lstinline{$fork} that provide an
interface to threads and need to be handled. To run the actor
\lstinline{$p}, we first need to pass it to the \lstinline{$act}
handler. This needs to be converted to a continuation so that it can
be further passed to the \lstinline{$scheduler} handler as defined in
Section~\ref{sec:dynamic-threads}.

\begin{lstlisting}
  (func $run_actor (param $p (ref $cont))
     (cont.new $cont_cont  (ref.func $act))  ;; convert the $act function into a continuation
     (cont.bind $cont_cont $cont  (local.get $p))  ;; partially apply the continuation to actor $a
     (call $scheduler) ) ;; handle fork & yield in the applied continuation
\end{lstlisting}

\section{Implementation}
\label{sec:implementation}
We have implemented the full instruction set of WasmFX as an extension
to the Wasm reference interpreter. Moreover, we have implemented a
prototype of WasmFX in Wasmtime, a production-grade Wasm engine. In
this section we describe the latter implementation.
Wasmtime features an optimising just-in-time compiler for Wasm
modules. We classify our implementation in Wasmtime as a prototype for
three reasons. Firstly, at the time of writing we support only WasmFX
programs on x86-64, whereas Wasmtime supports a much wider range of
architectures including ARM64, ARM, RISC-V64, s390x, and x86.
Secondly, we cannot implement $\RESUMETHROW$ as Wasmtime does not yet
support exceptions, thus we currently support only the other four
instructions.  We plan to implement $\RESUMETHROW$ once support for
exceptions lands in Wasmtime.
Thirdly, and most importantly, our implementation piggybacks on the
existing fiber API in Wasmtime~\cite{Crichton21}, which enables the
Wasm host to run functions asynchronously. Targeting the fiber API
allows us to quickly prototype the design in a production-grade
setting, and its implementation integrates well with standard
debugging and profiling tools as it already emits the necessary
information to construct DWARF unwind tables.  On the other hand, it
has the problem that fibers exist \emph{outside} the Wasm world,
meaning that every interaction with a fiber necessarily needs to call
out to the host, a relatively expensive operation.  Moreover, it
requires us to \emph{box} continuation parameters and results, because
the host function cannot be polymorphic over continuation types.

\topic{The Fiber API}
Figure~\ref{fig:fiber-api} shows the interface to the Wasmtime fiber
API. The \lstinline[style=rust]$FiberStack::new$ method allocates a
new stack of minimum \lstinline[style=rust]$size$
bytes. Figure~\ref{fig:stack-layout} details the actual stack
layout. The first 32 bytes are reserved for the header. The initial 8
bytes are used to store the payload for resumes, suspends, and
returns. The next 8 bytes are used for the local stack pointer. The
next 8 bytes store the pointer to the parent fiber. These bytes were
added by us to implement the dynamic scoping of handlers. Finally,
there is a guard page to detect stack overflow. We also extend the
instance context of Wasmtime with an additional field to keep track of
the currently executing fiber such that we can retrieve its pointer
when resuming and suspending.

The key entity is the \lstinline[style=rust]$Fiber$ structure, which
is parameterised by three type variables for resume payloads
(\lstinline[style=rust]$R$), suspend payloads
(\lstinline[style=rust]$S$), and return values
(\lstinline[style=rust]$A$). We instantiate \lstinline[style=rust]$R$
and \lstinline[style=rust]$A$ to the unit type, because we store
payloads directly in the bespoke buffer on the fiber stack. The reason
for using this approach is that we do not a priori which types the
user provided program will use. Consequently, we are forced to box
everything that crosses the Wasm-Host boundary and vice versa.
Meanwhile, we instantiate \lstinline[style=rust]$S$ to
\lstinline[style=rust]$u32$ as we use it to communicate index of the
control tag supplied to \SUSPEND.


\lstinline[style=rust]$Fiber::new$ takes a fresh
\lstinline[style=rust]$FiberStack$, and the suspended computation to
run on the said stack. It installs the launchpad code necessary to run
the computation when the initial \lstinline[style=rust]$Fiber::resume$
occurs. The \lstinline[style=rust]$Fiber::resume$ method suspends the
currently executing fiber and continues execution of the provided
\lstinline[style=rust]$Fiber$. The
\lstinline[style=rust]$Suspend::suspend$ method works similarly.

\topic{Stack Switching} Stack switching in the fibers library is
implemented in x86\_64 assembly code. Figure~\ref{fig:switchasm} shows
the stack switching routine. It pessimistically saves any necessary
registers to the stack. The pointer to the stack is passed in register
\lstinline$rdi$, from which we load the saved \lstinline$rsp$ (local
stack pointer); subsequently we store the current stack pointer in the
same header spot. The actual stack swapping occurs by overriding the
value of \lstinline$rsp$ with the previously loaded value. After
restoring saved registers, the top of the stack holds the return
location saved when the fiber was suspended; a return instruction is
executed, continuing execution from the suspended point.

\topic{Compiling WasmFX to Wasmtime Fiber}
The translation from WasmFX continuation instructions to Wasmtime
fibers is mostly straightforward. We implement each instruction as a
\emph{libcall} which calls from compiled Wasm code to Wasmtime runtime
code.  We realise each instruction as follows.

We map $\CONTNEW$ to \lstinline[style=rust]$Fiber::new$. We box the
resulting fiber so we can null out the continuation object once it has
been supplied to $\RESUME$. The suspended computation merely marshals
values to the provided Wasm function.
For $\CONTBIND$ we write the payload directly to the heap-allocated
result cell and return a new continuation object (containing the same
fiber reference).

The handler clauses on the $\RESUME$ instruction get compiled to the
core Wasm instruction \lstinline$br_table$, which is a jump table. The
table jumps to handler blocks for tags that are handled, and
re-suspends to the parent (if any) for tags that are not. The libcall
for $\RESUME$ boxes the arguments, sets the parent pointer of the
fiber to the currently executing fiber, and invokes
\lstinline[style=rust]$Fiber::resume$. We return the result of this
function to the handler in Wasm, distinguishing return from suspend
with a sentinel value.

The libcall for $\SUSPEND$ replaces the current stack pointer in the
context with its parent.  It then writes the payload, including the
name of the tag, to its result cell and transfers control.

\begin{figure}
  \centering \begin{minipage}[b]{0.47\linewidth} \centering
\begin{lstlisting}[style=rust]
trait FiberStack {
  fn new(size: usize) -> io::Result<Self>
}
trait<R, S, A> Fiber<R, S, A> {
  fn new(stack: FiberStack,
               func: FnOnce(R, &S<R, S, A>) -> A
  fn resume(&self, val: R) -> Result<A, S>
}
trait Suspend<R, S, A> {
  fn suspend(&self, S) -> R
}
\end{lstlisting}
  \subcaption{The Essence of the Wasmtime Fiber API}
  \label{fig:fiber-api}
  \end{minipage}
  \begin{minipage}[b]{0.45\linewidth}
    \centering
    \begin{lstlisting}[style=assembly]
.wasmtime_fiber_switch:
  // Save callee-saved registers
  push ...
  // Load resume pointer, save previous
  mov rax, -0x20[rdi]
  mov -0x20[rdi], rsp
  // Swap stacks
  mov rsp, rax
  pop ... // restore callee-saved registers
  ret
\end{lstlisting}
    \subcaption{Fiber Switching x86\_64 Assembly Code}
    \label{fig:switchasm}
  \end{minipage}
  \caption{Wasmtime Fiber API and Context Switching Code}
\end{figure}
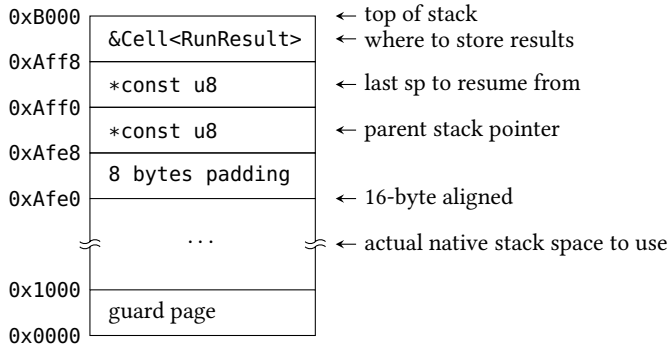
\begin{figure}[t]
  \pgfdeclaredecoration{discontinuity}{start}{
    \state{start}[width=0.5\pgfdecoratedinputsegmentremainingdistance-0.5\pgfdecorationsegmentlength,next state=first wave]
    {}
    \state{first wave}[width=\pgfdecorationsegmentlength, next state=second wave]
    {
      \pgfpathlineto{\pgfpointorigin}
      \pgfpathmoveto{\pgfqpoint{0pt}{\pgfdecorationsegmentamplitude}}
      \pgfpathcurveto
          {\pgfpoint{-0.25*\pgfmetadecorationsegmentlength}{0.75\pgfdecorationsegmentamplitude}}
          {\pgfpoint{-0.25*\pgfmetadecorationsegmentlength}{0.25\pgfdecorationsegmentamplitude}}
          {\pgfpoint{0pt}{0pt}}
      \pgfpathcurveto
          {\pgfpoint{0.25*\pgfmetadecorationsegmentlength}{-0.25\pgfdecorationsegmentamplitude}}
          {\pgfpoint{0.25*\pgfmetadecorationsegmentlength}{-0.75\pgfdecorationsegmentamplitude}}
          {\pgfpoint{0pt}{-\pgfdecorationsegmentamplitude}}
  }
  \state{second wave}[width=0pt, next state=do nothing]
    {
      \pgfpathmoveto{\pgfqpoint{0pt}{\pgfdecorationsegmentamplitude}}
      \pgfpathcurveto
          {\pgfpoint{-0.25*\pgfmetadecorationsegmentlength}{0.75\pgfdecorationsegmentamplitude}}
          {\pgfpoint{-0.25*\pgfmetadecorationsegmentlength}{0.25\pgfdecorationsegmentamplitude}}
          {\pgfpoint{0pt}{0pt}}
      \pgfpathcurveto
          {\pgfpoint{0.25*\pgfmetadecorationsegmentlength}{-0.25\pgfdecorationsegmentamplitude}}
          {\pgfpoint{0.25*\pgfmetadecorationsegmentlength}{-0.75\pgfdecorationsegmentamplitude}}
          {\pgfpoint{0pt}{-\pgfdecorationsegmentamplitude}}
      \pgfpathmoveto{\pgfpointorigin}
  }
    \state{do nothing}[width=\pgfdecorationsegmentlength,next state=do nothing]{
      \pgfpathlineto{\pgfpointdecoratedinputsegmentlast}
    }
    \state{final}
    {
      \pgfpathlineto{\pgfpointdecoratedpathlast}
    }
  }
  \newcommand{\width}{3cm}
  \newcommand{\height}{0.6cm}
  \newcommand{\arrowsep}{2ex}
  \newcommand{\contentsep}{1ex}
  \newcommand{\arrowlen}{2ex}
  \newcommand{\wigglesize}{1ex}
  \small
  \begin{tikzpicture}
    \draw (0, 0) node[left] {\texttt{0x0000}};
    \draw (0, 0) rectangle +(\width, \height);
    \draw (\contentsep, 0.5 * \height) node[right] {guard page};

    \draw (0, \height) node[left] {\texttt{0x1000}};
    \draw [decoration={%
        discontinuity,
        amplitude=\wigglesize, segment length=0.5 * \wigglesize, meta-segment length=\wigglesize},
    decorate] (0, \height) -- +(0, 2 * \height);
    \draw [decoration={%
        discontinuity,
        amplitude=\wigglesize, segment length=0.5 * \wigglesize, meta-segment length=\wigglesize},
    decorate] (\width, \height) -- +(0, 2 * \height);
    \draw (0.5 * \width, 2 * \height) node {$\cdots$};
    \draw [stealth-](\width + \arrowsep, 2 * \height) -- +(\arrowlen, 0) node[right] {actual native stack space to use};

    \draw (0, 3 * \height) node[left] {\texttt{0xAfe0}};
    \draw (0, 3 * \height) rectangle +(\width, \height);
    \draw (\contentsep, 3.5 * \height) node[right] {\texttt{8 bytes padding}};
    \draw [stealth-](\width + \arrowsep, 3 * \height) -- +(\arrowlen, 0)  node[right] {16-byte aligned};

    \draw (0, 4 * \height) node[left] {\texttt{0xAfe8}};
    \draw (0, 4 * \height) rectangle +(\width, \height);
    \draw (\contentsep, 4.5 * \height) node[right] {\texttt{*const u8}};
    \draw [stealth-](\width + \arrowsep, 4.5 * \height) -- +(\arrowlen, 0) node[right] {parent stack pointer};

    \draw (0, 5 * \height) node[left] {\texttt{0xAff0}};
    \draw (0, 5 * \height) rectangle +(\width, \height);
    \draw (\contentsep, 5.5 * \height) node[right] {\texttt{*const u8}};
    \draw [stealth-](\width + \arrowsep, 5.5 * \height) -- +(\arrowlen, 0) node[right] {last sp to resume from};

    \draw (0, 6 * \height) node[left] {\texttt{0xAff8}};
    \draw (0, 6 * \height) rectangle +(\width, \height);
    \draw (\contentsep, 6.5 * \height) node[right] {\texttt{\&Cell<RunResult>}};
    \draw [stealth-](\width + \arrowsep, 6.5 * \height) -- +(\arrowlen, 0) node[right] {where to store results};

    \draw (0, 7 * \height) node[left] {\texttt{0xB000}};
    \draw [stealth-](\width + \arrowsep, 7 * \height) -- +(\arrowlen, 0) node[right] {top of stack};
  \end{tikzpicture}
\caption{Fiber Stack Layout}
\label{fig:stack-layout}
\end{figure}

\subsection{Experiments}
\label{sec:evaluation}
We perform some preliminary experiments to gather data to obtain
insight into our prototype implementation of WasmFX fares against the
state-of-the-art.
We perform two experiments: 1) we measure the performance
characteristics, i.e. binary size, run time, and memory performance on
a micro benchmark provided by \citet{libmprompt}; 2) we compare the
binary size of programs produced by the TinyGo compiler with and
without WasmFX as a backend.
The experiments are conducted on an AMD Ryzen 9 5900X 12-core 3.75GHz
CPU with 32 GB memory powered machine running Ubuntu 22.04 LTS.

The first benchmark, provided by \citet{libmprompt}, simulates a web
server workload. It is written in C. The benchmark runs 10000
coroutines concurrently (each intended to represent a unique http
request), each coroutine simply suspends to simulate I/O and then
performs some stack-heavy computation. The benchmark performs 10
million requests, meaning it spawns 10 million coroutines in total.

We run the benchmark in three ways: 1) using WasmFX; 2) Asyncify; and
3) a bespoke hand-written state machine transformation of the
benchmark program. It is worth noting that the WasmFX and Asyncify
approaches require no programmer intervention, whereas the latter
approach requires a complete rewrite of the program. We run each
benchmark program five times and measure the size of the compiled Wasm
binary, median wall clock time, and maximum physical memory usage.
We use Wasmtime to run each benchmark. To compile the C program to
Wasm we use \texttt{clang} version 14 with optimisation flag
\texttt{-O3} and the binaryen toolchain version 105.
For the experiments, we modify the Wasmtime implementation of WasmFX
to allocate fiber stacks using malloc instead of mmap as this is the
memory allocation scheme used by the Asyncify and hand-written
program. In particular, we use the optimised \texttt{mimalloc}
allocator~\cite{LeijenZM19} rather than the default system
allocator. Finally, WasmFX and Asyncify allocate stacks with the fixed
size of 4096 bytes.

Figure~\ref{tbl:coroutine-bench} reports the results for the first
benchmark. The memory usage of WasmFX is close to that of the
state-of-the-art (optimised Asyncify). The WasmFX binary is
significantly more compact than the Asyncify binary. However, the run
time performance of the WasmFX implementation is more than a factor of
three slower. Performance analysis indicates that about 37\% of samples
are spent in \texttt{wasmtime\_fiber\_switch} (Figure
\ref{fig:switchasm}) with an additional 33\% of samples in the Rust
libcall code supporting the continuation instructions, which are both
highly unoptimised. Saving registers, calling between the host and
Wasm, and boxing values likely accounts for much of this cost. As
expected the bespoke implementation is fastest and exhibits the best
memory utilisation of all. The primary reason for the space efficiency
is that it does not allocate 4096 bytes stack for each coroutine.

\begin{figure}
  \centering
  \begin{minipage}{1.0\linewidth}
    \centering
  \begin{tabular}{rccc}
    \toprule
    & Binary Size & Wall Time & Memory Usage \\
    \midrule
    WasmFX & 0.8\,KB & 2700\,ms & 55.5\,MB \\
    Asyncify & 9.2\,KB & 700\,ms & 54.1\,MB \\
    Bespoke & 0.9\,KB & 140\,ms & 13.4\,MB \\ \bottomrule
  \end{tabular}
  \subcaption{Performance Results for WasmFX, Asyncify, and Bespoke}
  \label{tbl:coroutine-bench}
  \end{minipage}
  \begin{minipage}{1.0\linewidth}
    \centering
  \begin{tabular}{rccc}
    \toprule
    & main-kjp.go & coroutines.go \\
    \midrule
    WasmFX & 156\,KB & 7.2\,KB \\
    Asyncify & 597\,KB & 40\,KB \\ \bottomrule
  \end{tabular}
  \subcaption{Binary Size Comparison for TinyGo Programs}
  \label{tbl:tinygo-size}
  \end{minipage}
  \caption{Preliminary Results}
\end{figure}

In the second experiment we measure the potential effect that WasmFX
and Asyncify have on the Wasm binary size. Specifically, we compile
two off-the-shelf Go micro-benchmark programs, both of which makes
heavy use of coroutines~\cite{go-benchmark}. We compare the size of
the Wasm binary generated by the TinyGo compiler with WasmFX as a
backend and optimised Asyncify as a backend. For the WasmFX backend,
we use a coroutine scheduler which closely resembles that of
Section~\ref{sec:overview}.
The results are present in Figure~\ref{tbl:tinygo-size}. The compiled
size of the WasmFX powered binary is significantly smaller than the
state-of-the-art.

\section{Discussion}
\label{sec:discussion}

\subsection{Design Considerations}
\label{sec:design}

In this section we elaborate on the motivations for the design choices
of WasmFX.

\topic{Direct-style}
The goal of WasmFX is to provide a universal mechanism for
implementing non-local control features. So-called ``stackless''
approaches such as Asyncify~\citep{Zakai19} simulate such features on
top of Wasm by applying some form of global transformation into
continuation-passing style or a state machine~\cite{BiermanRMMT12}.
To support non-local control flow, functions must be instrumented,
resulting in both runtime cost and increase in code-size. The runtime
cost can be alleviated by compiling functions twice, once with support
for non-local control flow and once without, but that increases code
size further.  WasmFX provides non-local control natively in direct
style, providing a convenient target for features such as actors,
async/await, green threads, etc.

\topic{Delimited control}
It is well-established that delimited control operators provide a
universal mechanism for implementing non-local control features.
Effect handlers provide a structured form of delimited control
supporting fine-grained definition and modular composition of
non-local control features~\cite{KammarLO13}. Recent work has shown
that effect handlers can be implemented efficiently in
practice~\citep{SivaramakrishnanDWKJM21,GhicaLBP22}.
A key difference between effect handlers and traditional delimited
control operators is that the handling of a continuation occurs at the
delimiter rather than the capture-site of the continuation. In this
sense, effect handlers are similar to \citeauthor{Sitaram93}'s
run/fcontrol operator~\cite{Sitaram93}. Though, effect handlers
provide multiple typed control tags with implicit effect forwarding,
whereas run/fcontrol provides a single universal control tag, which in
turn means there is no notion of implicit effect forwarding. Instead
each instance of run is indexed such that an invocation of fcontrol
can be explicitly directed to a particular run instance (for more
details, the interested reader may consult \cite[Appendix
A]{Hillerstrom21} for a comprehensive survey of first-class control
operators).

\emph{Why not undelimited control?~}
It is possible to take undelimited control abstractions such as
call/cc~\citep{SperberDFvSFM09} or lightweight threads as
primitive. It is even possible to build general delimited control
abstractions, including effect handlers~\citep{GhicaLBP22}, on top of
these. However, undelimited continuations do not compose, which makes
working with undelimited control operators significantly more
difficult in practice~\citep{Kiselyov12}.
The composability of delimited control is what underpins the support
for structured concurrency features in Java 19, for
instance~\cite{Pressler18,BatemanP21}.

\topic{Synergy between the stack typing discipline of Wasm and effect handlers}
A further advantage of basing our design on effect handlers is that
the use of typed control tags fits seamlessly with the stack typing
discipline of Wasm. When suspending or resuming it is natural to
combine the associated context switch with the transfer of
data. Control tags allow for a choice of different kinds of suspension
each with their own type of payload which is passed unboxed on the
stack. Moreover, the result types of the control tag allow the
resuming of the suspension to pass back its own data unboxed on the
stack. The coupling of the parameter and result types of a control tag
provides a statically typed coupling between a suspension and its
corresponding resumption.

\topic{One-shot continuations support key use-cases and efficient implementation}
The WasmFX design naturally extends to support multi-shot
continuations, but the key use-cases we are targeting (primarily
different forms of concurrency) require only one-shot
continuations. One-shot continuations admit direct and efficient
implementations of continuations as stacks in which it is never
necessary to copy the stack. In principle, it would be possible to
extend Wasm with an affine type system to statically ensure that
continuations are invoked at most once, but this would place an undue
burden on producers and validators of Wasm code, so we doubt it would
be realistic. Instead, we follow the lead of OCaml 5 and dynamically
check that continuations are invoked at most once --- trapping in the
case that an attempt is made to invoke a continuation more than once.

\topic{Avoiding cycles}
Wasm currently includes no native support for automatic memory
management. There is a garbage collection proposal, but some producers
may prefer to not use it and some engines may not be able to support
it. WasmFX is consciously designed to avoid any dependency on garbage
collection. In particular, it supports an implementation that avoids
cycles in the heap and can thus be implemented using plain reference
counting. This property depends crucially on the affine semantics of
continuations: creating a cycle is possible only if a continuation can
get a handle to itself and stores it on its own stack.  However, a
continuation can only receive data when being resumed.  And once
resumed, a continuation is immediately marked dead.  An implementation
could hence null it out immediately, cutting the reference from
continuation to stack.  Consequently, even if a continuation is passed
to itself as a reference upon resumption, it will only receive a dead
reference that no longer points anywhere, preventing the possibility
of forming a cycle.

\topic{Avoiding allocation}
A downside of the implementation that avoids cycles is that it relies
on allocating a fresh continuation object every time we
\lstinline|suspend| or partially apply a continuation with
\lstinline|cont.bind|.  Perhaps, counter-intuitively, an
implementation --- e.g., one with GC --- can trade the no-cycles
property for an allocation-free implementation, where
\lstinline|suspend| and \lstinline|cont.bind| can reuse the same
physical continuation object each time, or in fact, merge it with the
stack itself.  The problems with doing this naively are twofold: 1)
each \lstinline|suspend| or \lstinline|cont.bind| can change the type
of the underlying continuation, and we must maintain type soundness;
2) we need to distinguish between these legitimate reuse and erroneous
attempts of using a continuation twice --- i.e., we need to be able to
distinguish live and dead references to the same physical continuation
object.

A solution to both these problems is to augment the representation of
continuation references and objects. We can represent a continuation
reference by a \emph{fat pointer} consisting of a pair of a pointer to
the continuation object and a unique sequence counter. The
continuation object itself also includes a sequence counter. Each time
it's consumed, the sequence counter of the object is incremented. If
there is a mismatch between the sequence counter of a continuation
reference and the object it points to, then this constitutes a
linearity violation and we trap. The sequence counter should at least
be a 64-bit integer as it is not unrealistic to expect a real program
to suspend more than $2^{32}$ times.  Under this implementation
approach, heap-stack cycles \emph{can} arise. Consider:
\begin{lstlisting}
(type $f (func (param (ref $c)))
(type $c (cont $f))
(tag $pause)
(func $task (param $x (ref $c)))
  (suspend $pause) )
(func $run
  (local $c (ref $c))
  (local.set $c  (cont.new $c  (ref.func $task)))
  (block $on_pause (result ...)  (resume (on $pause $on_pause)  (local.get $c)  (local.get $c)))
  ... );; here we get a contin. object that physically is the same as $c AND whose stack has a pointer to $c
\end{lstlisting}
If, on the other hand, \lstinline|suspend| allocates a fresh
continuation object, and the previous one is nulled out, then no cycle
will exist.  WasmFX's design, thanks to linear continuations, gives
implementations a choice between these different trade-offs.  This
choice is semantically transparent to Wasm code.

Another, more disruptive design would involve separating out the
sequence counter as its own special capability object. This would have
the advantage of avoiding the need for fat pointers at the cost of
having to change the interface to our instructions.

\topic{Unityped delimited continuations}
The bespoke Wasmtime Fiber API can be cast as an instance of delimited
control. Coincidentally, it provides operators with the same names as
the core instructions of WasmFX: \lstinline|new|, \lstinline|resume|,
and \lstinline|suspend|, which also behave quite similarly. In essence
it provides a unityped form of delimited continuations, where the
payload types for suspending and resuming are fixed, so in order to
support different types of payload the payload cannot be unboxed on
the stack. Instead of using a tag to determine where to handle a
suspend, it must always be handled by the immediate parent
context. This means there is no need to build any kind of special
handler construct into the syntax. Because payload types are fixed,
the type of the current continuation object does not change each time
it is resumed. This means that it can be reused in a type safe way
without the need for fat pointers or a linearity check, at the cost of
having to box heterogeneous payloads and not being able to support an
implementation that rules out cycles.

\topic{Combining handling with resumption}
Traditional accounts of effect handlers decouple handling of
computations from resuming of continuations. The WasmFX design instead
couples handling with resumption offering several advantages: 1) it
allows a different handler to be installed each time a continuation is
resumed; 2) it allows for the handling component of the
\lstinline|resume| instruction to be rather concise as it is simply a
mapping from tags to labels --- the actual handler code is attached to
the block structure defined by the labels; 3) it makes implementation
of stack segments simpler and more uniform as handlers are now in
one-to-one correspondence with active continuations.  Deep
handlers~\citep{PlotkinP09} automatically rewrap the current handler
around the resumption. Shallow
handlers~\citep{KammarLO13,HillerstromL18} instead require the handler
to be reinstalled each time a continuation is resumed. Deep handlers
are easier to reason about and optimise, but shallow handlers are more
convenient in some cases. The handlers in WasmFX can be seen as a
hybrid of shallow and deep: ``sheep handlers''. As with deep handlers
the body of a continuation is guaranteed to be wrapped in some
handler. As with shallow handlers this handler can be changed each
time we resume.

\topic{Return clauses}
Traditional effect handlers include an explicit return clause which
allows the handler to perform some transformation on the final value
returned by a computation. Return clauses are directly inspired by and
offer similar advantages to the success continuation of Benton and
Kennedy's~\citep{BentonK01} exceptional syntax variation of exception
handlers. WasmFX primitives on the other hand are lower level. Because
there is not automatically a join point between the code following a
\lstinline|resume| and the code following a handler clause it is easy
to wire in code to simulate a return clause. When we do want to join
up the control flow then we do so explicitly with a function call or a
branch instruction.

Consider a basic deep effect handler for a computation that returns an
integer which handles an operation \lstinline|ask| returning an
integer and has a return clause that converts the final result into a
floating point number.
\[
\ba{@{}l@{~}c@{~}l@{}}
\mathbf{return}~x &\mapsto& \textit{intToFloat}~x \\
\mathsf{ask}~k    &\mapsto& k~42 \\
\ea
\]
We might implement this in WasmFX as follows
\begin{lstlisting}
(type $icont (cont (func (result i32)))
(tag $ask (result i32))
(block $h  (result f32)
   (loop $l
      (block $on_ask (result i32 (ref $icont))
         (resume $cont (on $ask $on_ask)  (local.get $k))
         (f32.convert_i32_s) ;; return clause
         (br $h)
      ) ;; on_ask
      (cont.bind $icont $cont  (i32.const 42))
      (local.set $k)
      (br $l) ) )
\end{lstlisting}
where \lstinline|$k| is initialised with the computation to be
handled. Resuming the continuation leaves an \lstinline|i32| value on
the stack which the image of the return clause
\lstinline|(f32.convert_i32_s)| converts into a \lstinline|f32|.

 Its worth noting that in the typing rule for \lstinline|resume|
 (Figure~\ref{fig:typing-instructions}) the types $t_2^\ast$ always
 represent the original result types of the computation being handled
 and not the final result type after handling.

In the case of a scheduler, for instance, the code that dequeues the
next continuation and runs it constitutes the return clause. Though in
that case the return clause is fixed, because WasmFX handlers are
sheep handlers the return clause can change each time a continuation
is resumed. Its worth noting that in the typing rule for
\lstinline|resume| (Figure~\ref{fig:typing-instructions}) the types
$t_2^\ast$ always represent the original result types of the
computation being handled and not the final result type after
handling.

\subsection{Extensions}
\label{sec:extensions}

We strived to start with a minimal design for WasmFX, but of course,
many extensions are possible.

\newcommand\SUSPENDTO{\KK{suspend\_to}}
\newcommand\RESUMEWITH{\KK{resume\_with}}
\newcommand\HANDLERT{\K{handler}}
\newcommand\handlertype{\X{ht}}

\topic{Named handlers}
The core WasmFX design captures a continuation by suspending as far as
the nearest enclosing handler that matches the specified tag. An
alternative is to suspend to a specific named handler. We can support
such a mechanism by adding a new reference type $\HANDLERT\, t^\ast$
along with special variants of \lstinline|resume| and
\lstinline|suspend|.
The \lstinline|resume_with| instruction is just like
\lstinline|resume| except that it passes a fresh handler name to the
continuation.

\[
   \infer{
     \functype = ~t_1^\ast~(\REF\,\HANDLERT\, t_2^\ast) \to t_2^\ast
     \and
     (C \vdashhandler \handler : t_2^\ast \okhandler)^\ast
   }{
     C \vdashinstr \RESUMEWITH~\handler^\ast : t_1^\ast~(\REF\,\CONT \functype) \to t_2^\ast
   }
\]

The \lstinline|suspend_to| instruction is just like
\lstinline|suspend| except it takes an additional handler argument.

\[
  \infer{
    C_\CTAG(x) = t_1^\ast \to t_2^\ast
  }{
    C \vdashinstr \SUSPENDTO~\tagidx : t_1^\ast~(\REF\,\handlertype) \to t_2^\ast
  }
\]

A handler reference is similar to a prompt in a system of multi-prompt
delimited continuations~\citep{Gunter:mprompt}. However, since it is
created fresh for each handler, multiple activations of the same
prompt cannot exist by construction. The ergonomic tradeoffs between
named and unnamed handlers are reasonably apparent in high-level
source languages, but less so for a low-level target language like
Wasm. Unnamed handlers incorporate a form of dynamic binding, enabling
lightweight composition of effects. A natural way to simulate this
dynamic binding involves threading concrete handler implementations
through a program, something which can be painful to do manually,
impeding modularity. Named handlers offer a form of generativity
supporting a form of effect encapsulation~\citep{GhicaLBP22}. Unnamed
and named handlers, like single-prompt and multi-prompt delimited
control, can simulate one another (at a cost). In the future we plan
to measure that cost in Wasm. We suspect it may ultimately be
worthwhile to support both as some higher-level systems
do~\citep{GhicaLBP22,Xie:named}.

\topic{Barriers}
In some situations --- for example, when interacting with legacy code
that cannot or does not expect to be suspended and have interleaved
execution --- it is desirable to prevent suspensions beyond a certain
extent.  For this purpose, a block-like instruction $\BARRIER\,e^\ast$
can easily be introduced that bars any suspension from inside $e^\ast$
across its boundary.  A barrier may simply be viewed as a
``catch-all'' handler that handles any control tag by immediately
trapping.

\topic{Multi-shot}
Continuations in WasmFX are one-shot. Some applications of effect
handlers such as backtracking, probabilistic programming, and process
duplication exploit multi-shot continuations, but the key use-cases we
have in mind do not, and restricting attention to one-shot
continuations allows us to avoid having to copy stacks. Nevertheless,
it is natural to envisage a future extension that includes support for
multi-shot continuations by way of a continuation clone instruction.
However, some Wasm engines would have a hard time with such an
extension, since they use heterogeneous stacks mixing Wasm with C++
frames, which cannot easily be moved or copied.

\topic{Tail-resumptive handlers}
A handler is said to be tail-resumptive if it invokes the continuation
in tail-position in every handler clause~\cite{evidence_passing}. The
canonical example of a tail-resumptive handler is dynamic binding. The
handler clauses of a tail-resumptive handler can be inlined at the
suspend sites, because they do not perform any non-trivial control
flow manipulation, they simply retrieve a value. Inlining clauses
means that no time is spent constructing continuation objects. WasmFX
as it stands includes no facilities for identifying and inlining
tail-resumptive handlers. Moreover, the primary motivation for the
design for WasmFX is to support use-cases in which continuations are
\emph{not} invoked immediately. Nevertheless, it is natural to
envisage a future iteration of this proposal that includes an
extension for distinguishing tail-resumptive handlers.

\subsection{Related Work}
\label{sec:related-work}

\topic{Delimited control in Wasm} Wasm/k~\cite{PinckneyGB20} was an
early attempt to add first-class continuations to WebAssembly
1.0. Unlike WasmFX, it does not support multiple named control tags,
which are necessary to support several distinct uses of non-local
control in a typed and modular way.
Moreover, since Wasm/k did not consider emerging features of
WebAssembly 2.0, it does not compose with features that are now part
of the standard, such as typed function references, and others that
are in the process of being standardised, such as exceptions.

\topic{One-shot continuations} Using the call stack for implementing
one-shot continuations has a long history. Bruggeman et
al.~\cite{BruggemanWD96} show how to implement one-shot continuations
using segmented stacks in Scheme.  Farvardin et
al.~\cite{FarvardinR20} perform a comprehensive evaluation of various
implementation strategies on modern hardware including several
stack-based implementations of one-shot continuations such as
contiguous, segmented, and resizable stacks, as well as representing
continuations using a continuation-passing style transformation in the
compiler. The paper shows that if the primary concern is sequential
performance with advanced control-flow features, then contiguous or
resizable stacks are the best strategy.

\topic{Growing stacks} Resizeable segmented stacks are also used in the OCaml
implementation for delimited continuations~\cite{SivaramakrishnanDWKJM21}. The
OCaml managed stack starts out small but when the stack would overflow it is
reallocated -- potentially to a different location. This is safe in OCaml as
the compiler and runtime ensure that there are never any pointers into the
stack. This is generally not the case though in most languages, like C and C++,
and growing stacks by reallocation cannot be used in Wasm to implement effect
handlers.

The recent \emph{libmprompt} library~\cite{libmprompt} enables
segmented stacks at a system level where it can grow stack segments
\emph{in-place} by reserving virtual address space upfront, but
committing the memory on-demand when the stack would overflow. This
can be very efficient and could potentially be used for the WasmFX
implementation in Wasmtime.

\topic{Async/Await and Generators} Various languages, like C++,
Javascript, C\#, Rust, etc., implement specific instances of delimited
control in the form of async/await and generators.  To compile such
delimited control without runtime support (like WasmFX would
provide!), these languages require the async or generator functions to
be annotated, and then compile those functions in a special way that
allows them to be suspended on an await or yield. Generally each such
function allocates its stack frame in the heap, together with a state
machine that can resume at each await/yield
point~\cite{BiermanRMMT12}. It is possible to compile this efficiently
without allocation for small inlinable functions but with increased
nesting the overhead can be quite large compared to having runtime
support for direct stack switching. Moreover, due to the special
calling convention it is inherently not compositional where one needs
to decide upfront whether a function can be async or
not~\cite{whatcolor}.  \citet{evidence_passing} show how one can
compile general effect handlers using a monadic approach which avoids
the need for an explicit state machine and can use the standard C call
stack, where only on a suspend the stack is reified to an explicit
continuation.  This approach can be quite efficient but comes at the
cost of expanding the generated code at least by a factor 2 and, as
with async/await and generators, adds a complex compilation step.

\topic{Continuation marks} Chez Scheme supports continuation
marks~\cite{Flatt20}, a language feature that supports stack
inspection on top of which features such as exceptions, debuggers and
profilers are implemented in the presence of first-class
continuations. \citet{Flatt20} note that implementation strategy for
continuation marks and multi-prompt delimited continuations are
similar.

\subsection{Future Work}
\label{sec:future-work}

There are many opportunities for optimising the prototype
implementation. Here we outline some low-hanging fruit that we plan to
explore next. First, calling out to the Rust runtime code incurs an
overhead. Instead, the $\SUSPEND$ and $\RESUME$ instructions could be
compiled directly to architecture-specific instructions
(Figure~\ref{fig:switchasm}) which manipulate the registers, headers,
and stack pointer. Second, all values passed in $\RESUME$ and
$\SUSPEND$ are currently stored and loaded from memory. For
performance, these values could be passed directly, observing the
calling conventions of the host system. Finally, the fixed-sized
system stacks induce allocation burden that may be unnecessary:
experiments in other languages have found that most continuations
require little stack memory \citep{SivaramakrishnanDWKJM21}; other
allocation techniques such as those used by
libmprompt~\citep{libmprompt} may perform better.



We described a design and implementation for non-local control flow in
Wasm based on effect handlers. As such, WasmFX provides a unified and
composable extension that can directly express a wide variety of rich
control flow constructs found in various languages. We are looking
forward to strengthening our prototype implementation in
Wasmtime. Furthermore, we plan to add backends to various languages
with rich control flow to directly target the new WasmFX instructions.

\section*{Data-Availability Statement}
The software used to produce the results in
Section~\ref{sec:evaluation} is available on Zenodo~\cite{Artifact}.

\begin{acks}
  This work was supported by UKRI Future Leaders Fellowship ``Effect
  Handler Oriented Programming'' (reference number MR/T043830/1) and
  US NSF grant CCF-2102288.
  This material is based upon work supported by the Air Force Office
  of Scientific Research under awards number FA9550-17-1-0326 and
  FA9550-21-1-0024.
  We thank Alex Crichton and Nick Fitzgerald for helping us with the
  Wasmtime implementation.
\end{acks}

\bibliographystyle{ACM-Reference-Format}
\bibliography{\jobname}

\end{document}